\documentclass[final,5p,times,twocolumn]{elsarticle}

\usepackage{amssymb}
\usepackage{amsmath}

\usepackage{balance}

\usepackage{flushend}
 \usepackage{balance}
\usepackage{graphicx}
\usepackage{url}
\usepackage{listings}
\let\subparagraph\paragraph
\usepackage{fontawesome}

\usepackage[framemethod=TikZ]{mdframed}

\usepackage[hidelinks]{hyperref}

\newcommand{\tocheck}[1]{{#1}}

\definecolor{formalshadelight}{RGB}{242,242,242}
\definecolor{formalshadedark}{RGB}{166,166,166}

\newenvironment{highlight}[1][0]{

    \newcommand{\hnull}[2][\relax]{
        \begin{mdframed}[nobreak=true, hidealllines=true, leftmargin=0, rightmargin=0, innerleftmargin=1cm, innerrightmargin=0, skipabove=10pt, skipbelow=0, innertopmargin=5pt, innerbottommargin=0]
            \noindent%
            \ifx##1\relax
                \textbf{$\text{H}^\text{#1}_\text{0}$}: \textit{##2}\par
            \else 
                \textbf{$\text{H}^\text{#1.##1}_\text{0}$}: \textit{##2}\par
            \fi
        \end{mdframed}
    }

    \newcommand{\halt}[2][\relax]{
        \begin{mdframed}[nobreak=true, hidealllines=true, leftmargin=0, rightmargin=0, innerleftmargin=1cm, innerrightmargin=0, skipabove=0, skipbelow=0, innertopmargin=0, innerbottommargin=0]
            \noindent%
            \ifx##1\relax
                \textbf{$\text{H}^\text{#1}_\text{a}$}: \textit{##2}\par
            \else 
                \textbf{$\text{H}^\text{#1.##1}_\text{a}$}: \textit{##2}\par
            \fi
        \end{mdframed}
    }

    \vspace{5pt}
    \begin{mdframed}[nobreak=true, linewidth=1.75pt, linecolor=formalshadedark, topline=false, bottomline=false, rightline=false, backgroundcolor=formalshadelight, leftmargin=0pt, rightmargin=0pt, innerleftmargin=3pt, innerrightmargin=3pt]
}{
    \end{mdframed}
}

\makeatletter
\def\ps@pprintTitle{%
 \let\@oddhead\@empty
 \let\@evenhead\@empty
 \def\@oddfoot{\footnotesize\itshape
  Preprint accepted to Journal of Systems and Software\hfill}
 \let\@evenfoot\@oddfoot
}
\makeatother

\journal{Journal of Systems and Software}
\begin{document}

\begin{frontmatter}



\title{The Technical Debt Gamble: A Case Study on Technical Debt in a Large-Scale Industrial Microservice Architecture}




\author[wut_address]{Klara Borowa\corref{mycorrespondingauthor}}
\cortext[mycorrespondingauthor]{Corresponding author}
\ead{klara.borowa@pw.edu.pl}

\author[wut_address]{Andrzej Ratkowski}

\author[unifi_address]{Roberto Verdecchia}
\ead{roberto.verdecchia@unifi.it}

\address[wut_address]{Warsaw University of Technology, Institute of Control and Computation Engineering, Warsaw, Poland}

\address[unifi_address]{University of Florence, Software Technologies Laboratory, Italy}









\begin{abstract}

Microservice architectures provide an intuitive promise of high maintainability and evolvability due to loose coupling. However, these quality attributes are notably vulnerable to technical debt (TD). 
Few studies address TD in microservice systems, particularly on a large scale. This research explores how TD manifests in a large-scale microservice-based industrial system. 
The research is based on a mixed-method case study of a project including over 100 microservices and serving over 15k locations. Results are collected \textit{via} a quantitative method based static code analyzers combined with qualitative insights derived from a focus group discussion with the development team and a follow-up interview with the lead architect of the case study system. 
Results show that (1) \tocheck{simple static} source code analysis can be an efficient and effective entry point for holistic TD discovery, (2) inadequate communication significantly contributes to TD, (3) misalignment between architectural and organizational structures can exacerbate TD accumulation, (4) microservices can rapidly cycle through TD accumulation and resolution, a phenomenon referred to as ``microservice architecture technical debt gamble''. Finally, we identify a set of fitting strategies for TD management in microservice architectures.


\end{abstract}




\end{frontmatter}




\section{Introduction} 
Technical Debt (TD) is a metaphor that describes software-related constructs created with short-term gains in mind,  which may consequently impact the system's future maintenance and implementation of new functionalities~\cite{avgeriou2016managing}. 
In order to mitigate TD, architects may consider utilizing a microservice architecture (MSA), which, in contrast to monolithic architectures, provides the intuitive promise of simplified maintenance and evolution of systems due to the loose coupling between microservices, making it possible to change one microservice without affecting other microservices~\cite{Newman2015}~\cite{dragoni2017microservices}. 
However, this intuition is not fully supported by research. Maintainability research in MSA is lacking~\cite{li2021understanding} and as such, it is not clear how severely the additional coordination between microservices may hinder maintenance.
Additionally, microservice co-evolution~\cite{assunccao2023microservices} seems to be the practical norm instead of the separate evolution of microservices, which means that change in MSA systems may actually become extremely difficult.
As such, the impact of MSA on TD is not fully clear, and it clearly differs from monolithic systems. Thus, TD requires separate research in the context of MSA.

However, to date, only a small amount of research activities focused on TD in the context of MSA~\cite{villa2022systematic}, with most of the existing studies focusing on rather small systems, comprising microservices in the order of tens~\cite{verdecchia2024technical}. As a further research gap, many existing studies on TD in microservice-based systems focus on the migration between monolithic and microservice architectures~\cite{villa2022systematic}~\cite{lenarduzzi2020does} instead of systems originally conceived with a microservice architecture in mind. 
This research gap is of particular concern to the software engineering community since the topics of architectural technical debt are under-researched compared to code debt despite being a major issue for software systems~\cite{Avgeriou2023}. 

\tocheck{The aim of this research is to understand how TD manifests itself in a large microservice-based system that is used as case study for this research. In terms of research question (RQ), our inquiry aims to answer the following one:}

\noindent\tocheck{\textit{RQ: How is TD characterized in the industrial large-scale microservice-based system considered as case study?}}

\tocheck{With our research question (RQ), we aim to systematically achieve the goal set out by our industrial partner, namely to gain insights into the TD of their system. This goal needs to be achieved by utilizing the boundaries set out by our partner, namely a combination of simple static analysis tools already in place at the  organization and direct access to the developers of the software product, in order to identify and understand the technical debt present in the system.}

In order to answer our RQ, we performed an analysis of the Retail System (RS), a software-intensive system comprising over 100 microservices, which is currently used to operate a network of over 15k convenience stores. We based our analysis on both quantitative and qualitative methods, namely reverse engineering the microservices relationships and measuring their workload based on the Azure DevOps tools and static source code analysis using SonarQube, followed by a focus group discussion with a development team, and a closing interview with the lead architect of the system.


The main contributions of this research are: (1) a mixed methods case study of a large microservice-based industrial system in the context of the system's technical debt, (2) a showcase of how this in-depth analysis of TD with the use of \tocheck{simple code analysis}, as well as a focus group discussion and interview with the architect, were performed, (3) the finding that \tocheck{simple code analysis} can be a lightweight yet effective entry point for comprehensive TD discovery, (4) the identification of the main causes of TD in large-scale industrial case study system, (5) uncovering the ``microservice architecture technical debt gamble'' phenomenon characteristic of MSA, (6) the identification of a set of best practices for TD management in MSA, (7) additional material containing the \tocheck{simple code analysis} results, focus group discussion plan and slides~\cite{additional_material_jss}.

This paper is structured as follows. We present research related to this study in Section~\ref{sec:related_work}. In Section~\ref{sec:case_study_description} we provide details about the software-intensive system researched in this case study. The research process followed is detailed in Section~\ref{sec:research_process} and the results of this inquiry are presented in Section~\ref{sec:results}. The findings of this research are documented in Section \ref{sec:discussion}, while the related threats to validity are reported in Section \ref{sec:threats_to_validity}. The concluding remarks of this study are drawn in Section~\ref{sec:conclusion_and_future_work}.

\section{Related Work} 
\label{sec:related_work}

\subsection{Technical debt identification}

In this case study, we focused on inadvertent TD~\cite{Fowler2009}, which can be hard to identify and monitor. Often, it is only when negative TD symptoms arise, e.g., the system's performance significantly slows down, that any efforts of finding this TD are taken~\cite{kruchten2019managing}.
Such identification can be done using various approaches~\cite{alves2016identification}, most notably: the use of automatic TD detection tools~\cite{avgeriou2020overview}, human knowledge based analyses~\cite{verdecchia2018architectural}, and detection of anti-patterns~\cite{lahti2021experiences}.

While automatic TD detection tools are widely available and relatively easy to use, it is important to note their main drawbacks. Firstly, they usually measure TD through code smells~\cite{alves2016identification}, i.e., signs that something ``smells bad'' in the code~\cite{fowler2018refactoring}, which can be an indicator of TD but are not a definitive sign of it. Even if particular pieces of code are sub-optimal, they may not cause maintenance or evolution-related problems. Additionally, TD detection tools make use of very different algorithms and often give contradictory results~\cite{lefever2021lack}. Despite these drawbacks, tools detecting code TD, such as SonarQube, are widely used in industry~\cite{lenarduzzi2023critical}~\cite{avgeriou2020overview}. In the case of this study, our industry partner had created an environment that allowed us to run SonarQube, which they recognized as useful for detecting code-related technical debt in their organization.

Architectural technical debt (ATD) is a TD type that is of particular interest to this study since we focus on an MSA system. Although finding ATD is possible by identifying architectural antipatterns and smells~\cite{verdecchia2018architectural}, detecting TD in architectural structures is difficult to automatize~\cite{lefever2021lack}. Tools for ATD detection do exist~\cite{fontana2017arcan}~\cite{avgeriou2020overview}~\cite{ospina2021atdx} but are not as widely adopted as SonarQube~\cite{avgeriou2020overview}. In this study, instead of focusing on ATD detecting tools, we focused on finding whether this most popular TD identification tool could be used to identify which microservices are most impacted by technical debt. Therefore, in order to get an additional understanding of ATD in this case study, we complemented the SonarQube data gathering with a focus group discussion with the RS system's development team.

\subsection{Technical debt in microservice systems}

By focusing on the research considering TD in microservice architectures, we were able to pinpoint only a limited number of studies currently available in the literature. We present them in this section while showcasing how our study differs from the existing research.

The case study by Verdecchia et al.~\cite{verdecchia2024technical}, which also focuses on TD in a microservice system, closely resembles this study in terms of the research method used, by blending a mix of qualitative and quantitative approaches. In contrast to that study however, the case study of this inquiry is characterized by a much larger size, comprising 30 key microservices from a system of over 100 microservices rather than 13. Additionally, the case study presented in this paper focuses on understanding the current TD in a software system, rather than studying how TD evolves over time. Finally, Verdecchia et al. researched an open-source software project, rather than a large-scale industrial one.

The study by Lenarduzzi et al.~\cite{lenarduzzi2020does} also resembles this study from a methodological standpoint, by adopting a quantitative analysis followed by a complementary qualitative one. It differs from this research however, as Lenarduzzi et al. consider the impact on TD by migrating a monolithic architecture to a microservice one. The case study reported in this inquiry instead considers a system that was originally conceived as a microservice architecture. By focusing on the number of microservices, we also note that the system considered in this study presents a much higher number of microservices (30 key microservices rather than 5), and could hence be deemed of bigger size and potentially more complex in nature.

In another case study by Toledo et al.~\cite{de2021identifying} interviews are conducted to create a catalog of architectural TD, consequences, and solutions. Unlike this study, Toledo et al. focus their research process on a purely qualitative method, by conducting interviews with practitioners. From a more semantic stance, in this study we do not focus on a specific type of TD, while the study of Toledo et al. explicitly concentrates on ATD. 

By considering the more encompassing topic of software quality in microservice-based systems, further related studies can be identified.
Bogner et al.~\cite{bogner2019assuring} present a qualitative study on evolvability assurance. While in their study TD emerged as a topic related to microservice evolvability, unlike this work, the study by Bogner et al. does not focus on TD. Methodologically, the study also differs by utilizing semi-structured interviews rather than a mixed-method case study. 

In another work related to the maintainability of microservice architectures, Pagazzini et al.~\cite{pigazzini2020towards} discuss how microservice smells can be identified in software projects through source code analysis. A major difference with that work is that this study presents a mixed-method case study rather than using only a source code analysis tool, and focuses on TD rather than maintainability. 

By considering the more encompassing topic of technical debt in service-oriented architectures (SOA), Nikolaidis et al.~\cite{nikolaidis2022technical} document a methodology to measure TD in services by constructing a call graph. Differently, by following the directives of the industrial partner involved in this research, we measure TD \textit{via} the widely adopted SonarQube rather than the approach of Nikolaidis et al., and specifically consider a microservice-based architecture rather than a general SOA. 

Through a systematic mapping study on TD in microservice architectures, Villa et al.~\cite{villa2022systematic} present a literature review on technical debt in microservices. From the literature analysis, there are few studies leveraging a quantitative research component to study the topic result, laying a grounding motivation for the research method used in this study.

\section{Case Study Description}
\label{sec:case_study_description}

The case study focuses on a sales support system Retail System (RS) designed and implemented by a retail organization operating a large network of over 15k convenience stores. The main purpose of the system is to expose the capabilities of point-of-sale (POS) applications to electronic sales channels, e.g., end customers' mobile apps and external partners.

\begin{figure}
\center
\includegraphics[width=0.5\textwidth]{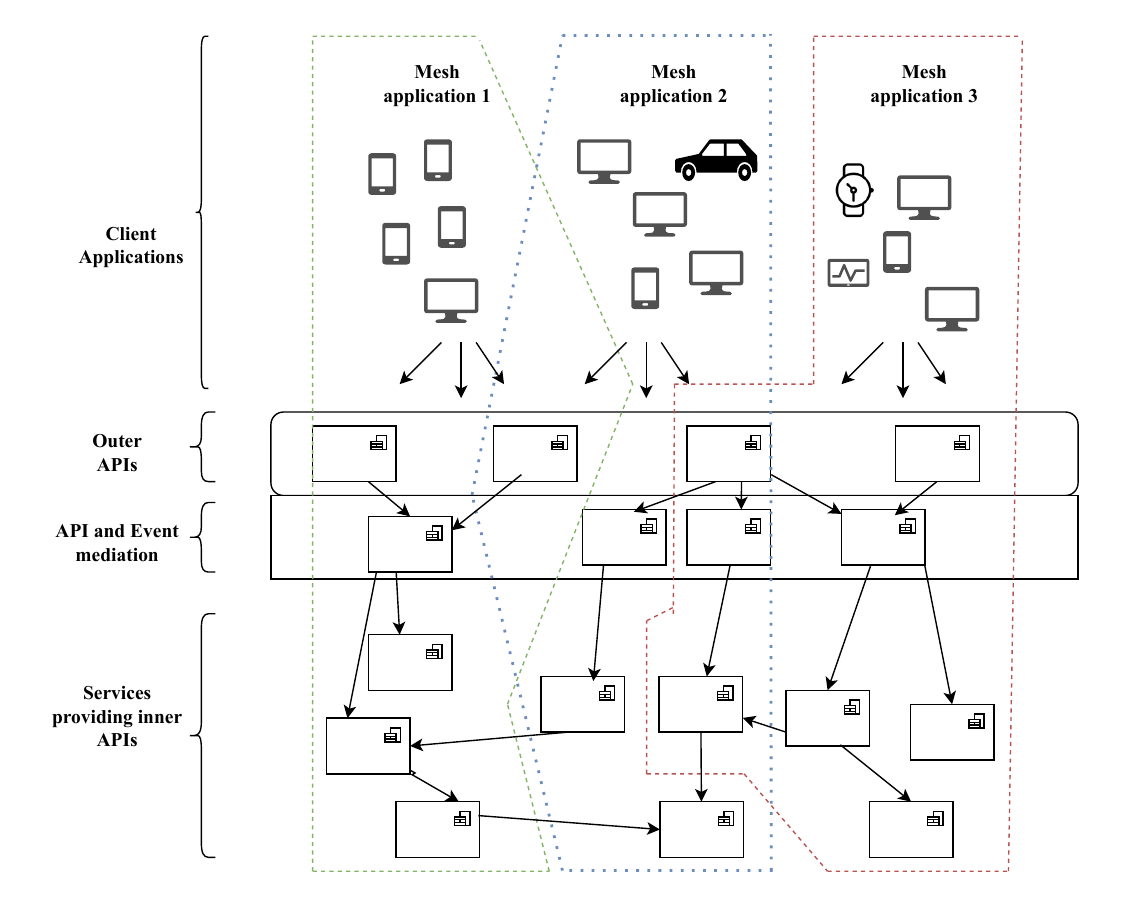}
\caption{Mesh App and Service Architecture (MASA) example (based on~\cite{gartner2019masa})} \label{fig:masa}
\end{figure}

The architecture of the RS is based on the Mesh App and Service Architecture (MASA)~\cite{gartner2019masa} framework.
MASA is a Gartner-proposed architectural framework for enterprise-size systems that are supposed to support divergent applications and business processes through multi-layer APIs. The key concept revolves around dividing the architecture into three layers (see Figure~\ref{fig:masa}): 

\begin{itemize}
\item  Outer APIs that serve external applications. 
\item  The API and Event Mediation layer that provides consistency in access management to inner APIs, reflecting business processes in the organization.
\item Services providing inner APIs that expose capabilities of enterprise systems (e.g., ERP, Point of Sale, appliances in shops).
\end{itemize}

Besides layers, MASA introduces meshes that crosscut vertically all three layers for the particular needs of specific client applications, e.g., a mesh for individual customers or a mesh for third-party partners.

In the RS, the architecture is spread across more than 100 microservices that are organized into three different layers in accordance with the MASA pattern. \tocheck{In the RS these layers are named slightly differently, so we present the corresponding MASA name in brackets. These layers are as follows:}

\begin{itemize}
\item \textbf{Experience API layer (in MASA: Other APIs layer)}: This layer constitutes the top external layer of the system, and groups the microservices used by the end-consumers. The microservices in this layer represent atomic behavior (also referred to as ``experience'') for a certain capability the application provides. Microservices generally possess simple functionality, e.g. requests authorization or data transformation between various formats. Microservices belonging to the experience layer do not contain, by design, any business logic.

\item \textbf{Orchestration layer (in MASA: API and Event mediation)}: This layer consists of microservices implementing the logic of the application business processes. Microservices in the orchestration layer are invoked by the microservices of the experience layer. 
Through each step of a specific business process, appropriate calls are made by microservices from the orchestration layer to services layer microservices.

\item \textbf{Services layer (in MASA: Services providing inner APIs)}: The bottom layer of the RS groups microservices that represent isolated functionalities used to perform the steps of various business processes. 
The Services layer microservices either use separate databases for storing their data or use interfaces (wrappers) to call external point-of-sale systems. 
\end{itemize}

\begin{figure*}
\includegraphics[width=\textwidth]{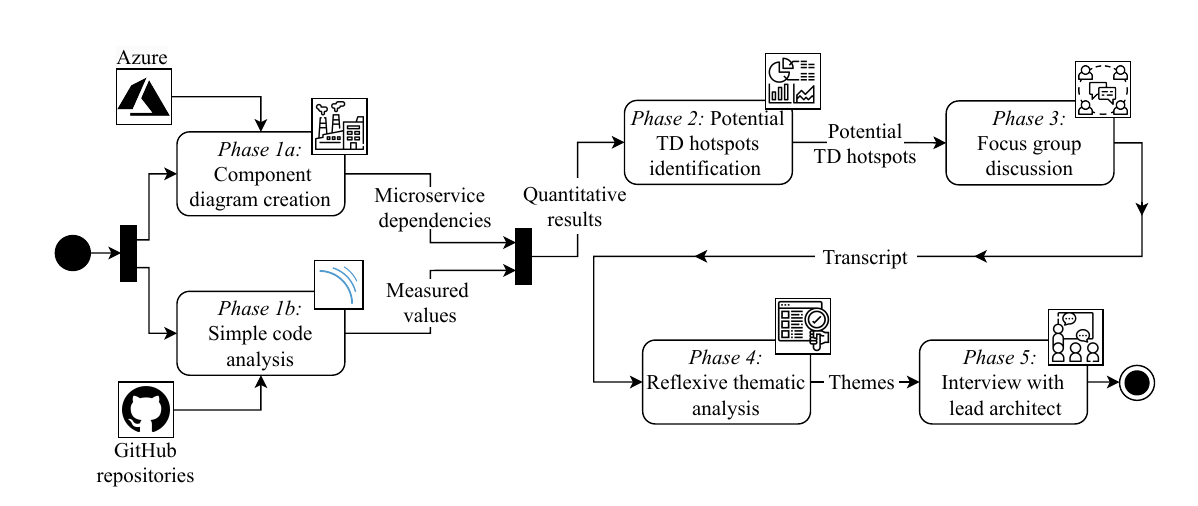}
\caption{Research process overview} \label{fig:method}
\end{figure*}

All microservices of the RS are deployed in a public cloud infrastructure. Microservices are developed and operated by multiple development teams located in various places around the world, belonging to different time zones, backgrounds, cultures, and work styles. While the whole system consists of over a hundred microservices, this case study focuses on the 30 core microservices of the RS, as selected by the lead architect of the system according to the interest of the industrial partner. The development and maintenance of the RS strongly \tocheck{employs a DevOps~approach, i.e. one development team is responsible not only for delivering new features of software but also for the stability and reliability of all software components used by their microservices. They have accountability for most tasks, such as maintenance, bug fixing, system monitoring, and operational efficiency (e.g. infrastructure cost tracking and optimization). Due to this fact, technical debt is a vital concern for the development team.}

The RS was a microservice-based system since its inception when the Enterprise Architecture team of the company decided to employ the MASA pattern. As such, the system was never a monolith or a distributed monolith, and all microservices can be changed and deployed separately.

Due to the nature of the RS business,  stakeholders require the system to be easily modifiable, maintainable, and evolvable, simultaneously providing high reliability, availability, and performance. The RS is fully operational and supports over 15k stores, millions of customers, and millions of sale transactions daily. 

\tocheck{The partner company has specifically requested to find TD that may impact the maintainability and evolvability of the RS system. As such, this study focuses on these two software quality attributes.}

\section{Research Process} 
\label{sec:research_process}
To systematically study the TD present in RS we adopted the mixed method empirical research process depicted in Figure~\ref{fig:method}.

The research process, consisting of five main phases, starts with an initial quantitative analysis which results in a component diagram and SonarQube analysis (Phases 1-2). The output of the quantitative analysis, namely a list of microservices that could be most affected by TD (also simply referred to as ``\textit{TD hotspots}'' in this paper), is then used as input for a qualitative analysis. During the qualitative analysis, a focus group with RS developers and an interview with the lead architect (Phases 3-5) were held to gain further insights, complement the quantitative results, and gain a comprehensive understanding of the TD in the RS.

We prepared material containing the full SonarQube results, the focus group discussion plan, and focus group discussion slides~\cite{additional_material_jss}. To preserve the anonymity of the partner company, the employees partaking in this study, and the RS, all presented data is anonymized by omitting personal names, component names, and other potentially sensitive information.


\subsection{Phase 1a: Component diagram creation.}
\label{method:diagram_creation}
The goal of this phase was to identify a set of microservices to be researched in the context of TD. We assumed that the most crucial microservices are the ones that process the most requests in the system and sought to identify them.

The first research step we performed was the identification of the core microservices used in the RS, as well as the relationships between them. This information was captured \textit{via} the Azure Application Insights \footnote{\url{https://learn.microsoft.com/en-us/azure/azure-monitor/app/app-insights-overview}. Accessed 4 April 2024} and the Azure Application Map distributed tracing system\footnote{\url{https://learn.microsoft.com/en-us/azure/azure-monitor/app/app-map}. Accessed 4 April 2024} based on the dynamic behavior of microservices at runtime. 
\tocheck{This analysis employed the following steps:}

\tocheck{
\textbf{Microservice instrumentation.}
Each microservice, running as java application, was equiped with the Application Insights telemetry instrumentation, which is responsible for gathering all generated microservice logs. The instrumentation is based on the "autoinstrumentation" approach~\cite{Azure2025_AutoInstr}. In practice, this means that each microservice, which was encapsulated in a Docker container, had a Dockerfile which contained an appropriate declaration as showcased in Listing~\ref{code:Dockerfile_instrument}.
}

\begin{lstlisting}[caption=Dockerfile fragment necessary for Azure instrumentation, label=code:Dockerfile_instrument, frame=single]
FROM gradle:6.9.4-jdk11 AS build
COPY --chown=gradle:gradle . /build/service

WORKDIR /build/service
RUN gradle build --no-daemon --console verbose test

RUN ls -la /build/service/build/libs
FROM adoptopenjdk:11-jre
WORKDIR /opt/service

COPY --from=build /build/service/build/libs/*.jar ./service.jar
ADD https://github.com/microsoft/ApplicationInsights-Java/releases/download/2.6.4/applicationinsights-agent-2.6.4.jar ./agent.jar

EXPOSE 8080
ENTRYPOINT ["/exec java -Djava.security.egd=file:/dev/./urandom \
   -XX:+UnlockExperimentalVMOptions \
   -XX:+UseShenandoahGC \
   -Xms1g \
   -Xmx6g \
   -javaagent:/opt/service/agent.jar \
   -jar /opt/service/service.jar"]    
\end{lstlisting}

\tocheck{
As a result of autoinstrumentation, all logs generated by applications run using the java virtual machine are transmitted into Azure Application Insights database.}

\tocheck{\textbf{Microservice relationship discovery.} On the basis of Azure Application Insights logs, it is possible to discover the relationships between all microservices in the system.} 

\tocheck{Each external API request to the RS system 
is automatically assigned a unique identifier known as operationId. This operationId is subsequently passed along with every subsequent call to the next microservice in the call chain sequence.}

\tocheck{Each request from this call chain sequence is logged in the Application Insight time-series database, facilitating efficient log querying. Beyond the operationId and request identifiers, each request is enriched with additional metadata, such as the microservice name and timestamp.}

\begin{figure}
\includegraphics[width=0.45\textwidth]{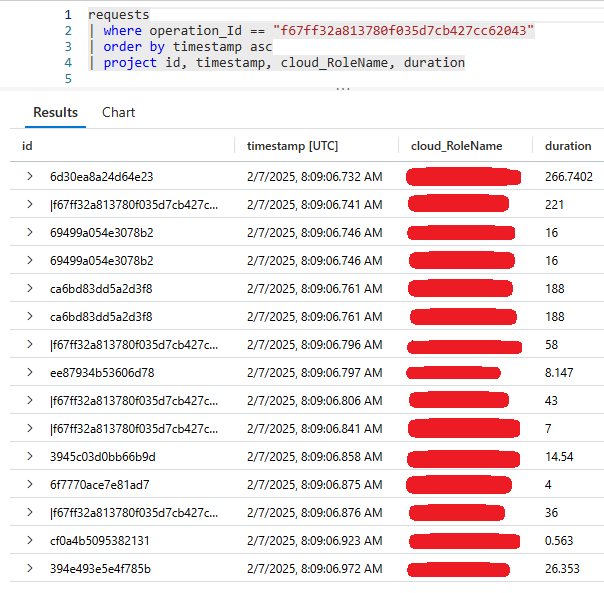}
\caption{Example query result from Application Insight database} \label{fig:ApplicationInsightQuery}
\end{figure}

\tocheck{To find the relationships between services, one can query the log database using a selected operationId and sort the results by the names of microservices in chronological order through a KQL (Kusto Query Language) query (see Figure~\ref{fig:ApplicationInsightQuery}). }

\begin{figure*}
\center
\includegraphics[width=\textwidth,scale=0.25]{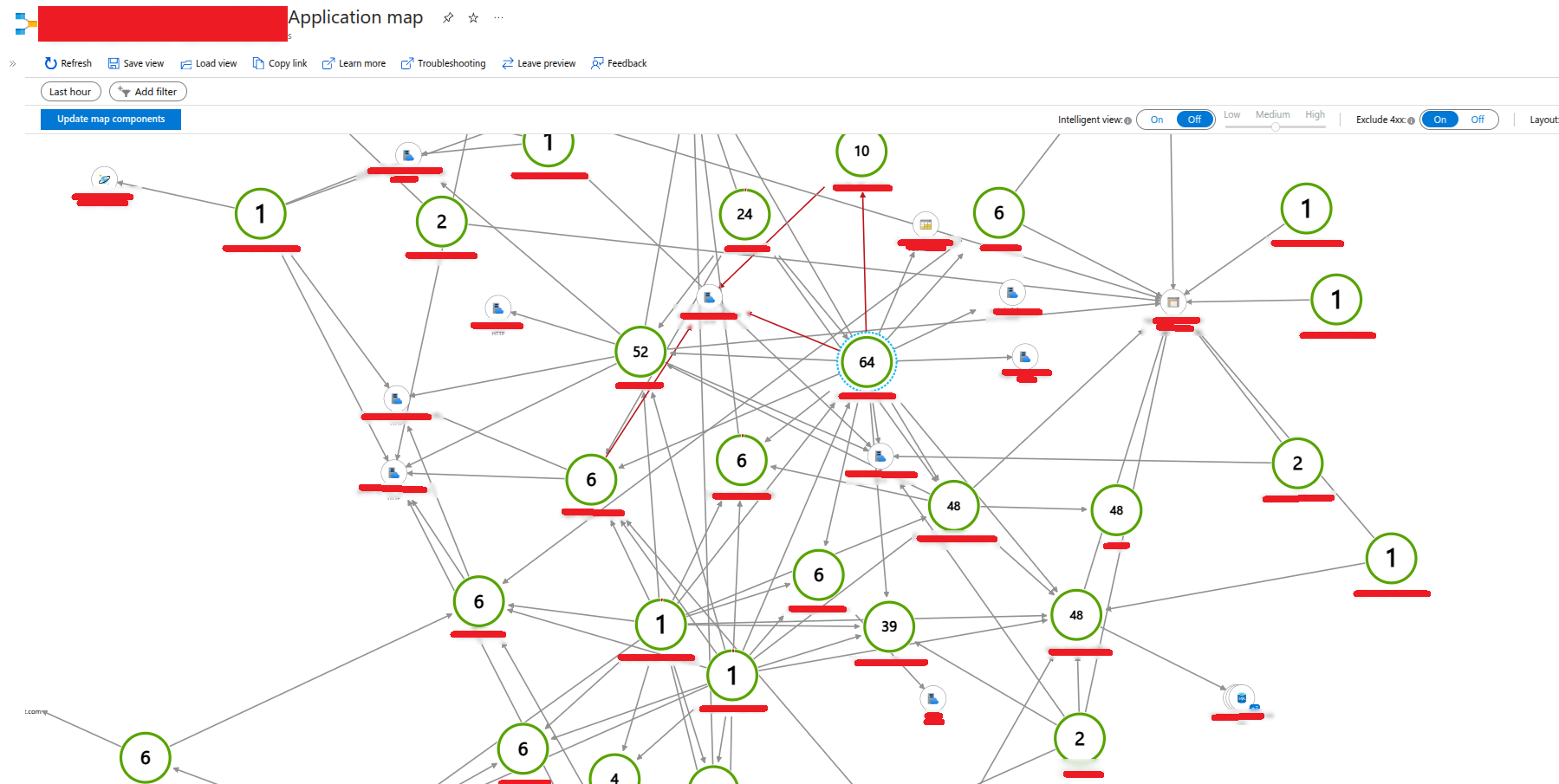}
\caption{Application Map showcasing a fragment of the RS architecture with anonymized microservice names} \label{fig:ApplicationMap}
\end{figure*}

\tocheck{The outcome of this query reveals the sequence of calls among a group of microservices, thereby illustrating the caller-callee relationships between them. This structured approach to logging makes it possible the find microservice interactions within the system's architecture. One specific Application Insight tool, the Application Map, produces a visualization of all relationships between microservices constructed from these logs on a single diagram (see Figure ~\ref{fig:ApplicationMap}).}

\tocheck{Thus, using the Application Map and KQL queries, we were able to obtain all relationships between microservices in the RS and the amount of requests sent and received by these microservices daily.}

\tocheck{\textbf{Microservices selection.}}
The choice of the 30 core microservices was made as follows. The lead architect used Azure Application Insights to gather information about the average daily number of requests handled by each microservice. This way, the top 20 microservices that handled most requests were identified as initial core microservices. Then, the architect found which microservices communicated with the initially selected 20 microservices and added them to the core microservices list, which resulted in 30 microservices in total.

The average number of daily requests the RS handles is 66.5M. Each of the identified 30 microservices handles, on average, from 17k to 15.1M requests daily, which means that they are overall responsible for over 99\% of the daily request load.
These microservices perform core business processes in the RS, such as sales, billing, warehouse management, and pricing.

The final output of this phase, namely the reverse-engineered microservice component diagram, is presented in Figure~\ref{fig:plant_uml}.

\tocheck{\subsection{Phase 1b: Simple Code Analysis}}
\label{method:sonarqube}

\tocheck{This phase's aim was to obtain an automatized quantitative analysis of the microservices' code, focusing on measurements that may possibly be connected with TD. }

\tocheck{As such, we gathered the following:
\begin{itemize}
    \item Number of dependencies: Since the complexity of a microservice could be related to the number of its dependencies. This information was gathered from the Azure Application Map.
    \item Lines of code number: As a microservice composed by a high number of lines of code may point to a lack of refactoring or overly complex code. This measurement was obtained straight from the microservices' GitHub repositories. 
    \item Age: Given that software aging naturally leads to software quality deterioration~\cite{lehman1979understanding}. The age was obtained from the GitHub repositories as well. 
    \item Number of contributors: Since human factors can have a major role in incurring TD~\cite{verdecchia2021building}, a high number of contributors could be a symptom (or cause) of a TD hotspot. The contributor count was gathered from the GitHub repositories.
    \item  Rarely adopted programming languages: Microservices written in a programming language different from all other microservices could indicate a TD hotspot, since developers may not be able to efficiently work with a rarely used technology. This information was also gathered from GitHub repositories.
    \item Technical Debt: The TD measurement directly represents the amount of time needed to repay the TD in minutes (calculated by SonarQube). This measurement was used to calculate TD density. 
    \item Technical Debt Density: Since a microservice with high TD density may be a major source of TD. Technical Debt Density represents the amount of time needed to repay the TD in minutes, divided by the lines of code number.
\end{itemize}}

\tocheck{The software department of the retail organization employs a multi-repository code-level strategy, where each microservice is encapsulated in a dedicated GitHub repository. Many of the measurements used in our analysis were obtained directly from these repositories (lines of code, age, contributors programming languages).}

\tocheck{The number of dependencies was obtained using the Azure Application Map, as described in Section~\ref{method:diagram_creation}.}

\tocheck{Additionally, in order to calculate TD repayment time, we decided on using an automatic TD detection tool instead of methods such as architectural antipattern analysis, since we knew of the use of the MASA~\cite{gartner2019masa} approach in the company and it seemed unlikely, due to strict company policy, for architectural antipatterns to emerged unnoticed. }

\tocheck{We chose SonarQube\footnote{\url{https://www.sonarsource.com/products/sonarqube}. Accessed 4 April 2024}, along with the other tools (Azure, GitHub), since the partner company itself has set up an environment for running SonarQube, which they used themselves in their delivery pipeline. The company itself got interested in using SonarQube~\cite{avgeriou2020overview} due to its popularity among TD measurement tools and, thus, was able to share this data with us without disrupting their daily operations. The partner company was particularly interested in whether SonarQube could be used to identify the most expensive TD in the system by employing SonarQube analysis in the process. }

\tocheck{We considered TD identification methods related to tracking changes in the repositories. However, we decided not to employ them for two reasons: (1) the lack of change in code can actually be a symptom of TD, since developers may become afraid of ``touching'' TD impacted code~\cite{verdecchia2021building}, (2) the partner company was insistent on using simpler code analysis instead of employing such tools.}

\tocheck{We adopted SonarQube v10.2.1 with SonarScanner v50.0.1 and the default TD repayment cost of 30 minutes per line of code.  As a result, we obtained a TD measurement in minutes, i.e., the amount of minutes necessary to repay the TD according to SonarQube. The Technical Debt Ratio was then calculated by dividing the TD measurement by the lines of code value obtained from the microservice's repository.}


\subsection{Phase 2: Potential TD hotspots identification.}
\label{method:potential_TD_identification}
In this phase, we sought to find potential TD hotspots, i.e., microservices that the quantitative analysis suggested could be most affected by TD.
\tocheck{In order to do so, we analyzed the data from Phase~1a and Phase~1b, while focusing on the following criteria: the highest number of dependencies, the highest number of lines of code, the biggest age, the highest number of contributors, use of unique programming languages not present in the rest of the system, the highest TD measurement, the highest TD density.}


To verify the meaningfulness of the TD hotspots identified according to the criteria listed above, we presented the preliminary list of TD hotspots to the lead architect of the RS. The lead architect supported the final selection of potential TD hotspots. In addition, from this preliminary consultation, we learned that one known TD hotspot was missing, namely a ``proof of concept that stayed''~\cite{verdecchia2021building} microservice, was missing from the list. Therefore, we included this additional microservice.

The output of this phase resulted in a list of eight microservices that were potentially the most impacted by TD. This list was used as the basis for discussion to kickstart the focus group, i.e., the first phase of the qualitative analysis (Phase 4).

\begin{table}
\scriptsize
\label{tab:participants}
\begin{tabular}
{llllll}
\hline
ID	&Education&ED-CS&Role&EXP&EXP-COMP\\
\hline
1	&	PhD	&Yes&	Architect	&	20	&	6	\\
2	&	Bachelor's&No&	Business\&System Analyst&	9	&	2.5	\\
3	&	Master's&No&	Quality Assurance	&	6	&	4	\\
4	&	Master's&No&	Scrum Master	&	8	&	1.5	\\
5	&	Master's&No&	Developer	&	5	&	2.5	\\
6	&	Master's&Yes&	Developer	&	11	&	2.5	\\
7	&	Master's&Yes&	Developer	&	11	&	2	\\
8	&	Master's&	Yes&Developer	&	5	&	1	\\

\hline
\end{tabular}
\caption{Focus group discussion participants \\ED-CS -- Education related to computer science, \\EXP -- Years of experience in software development, \\EXP-COMP -- Years in current company}
\end{table}

\begin{figure*}
\includegraphics[width=\textwidth]{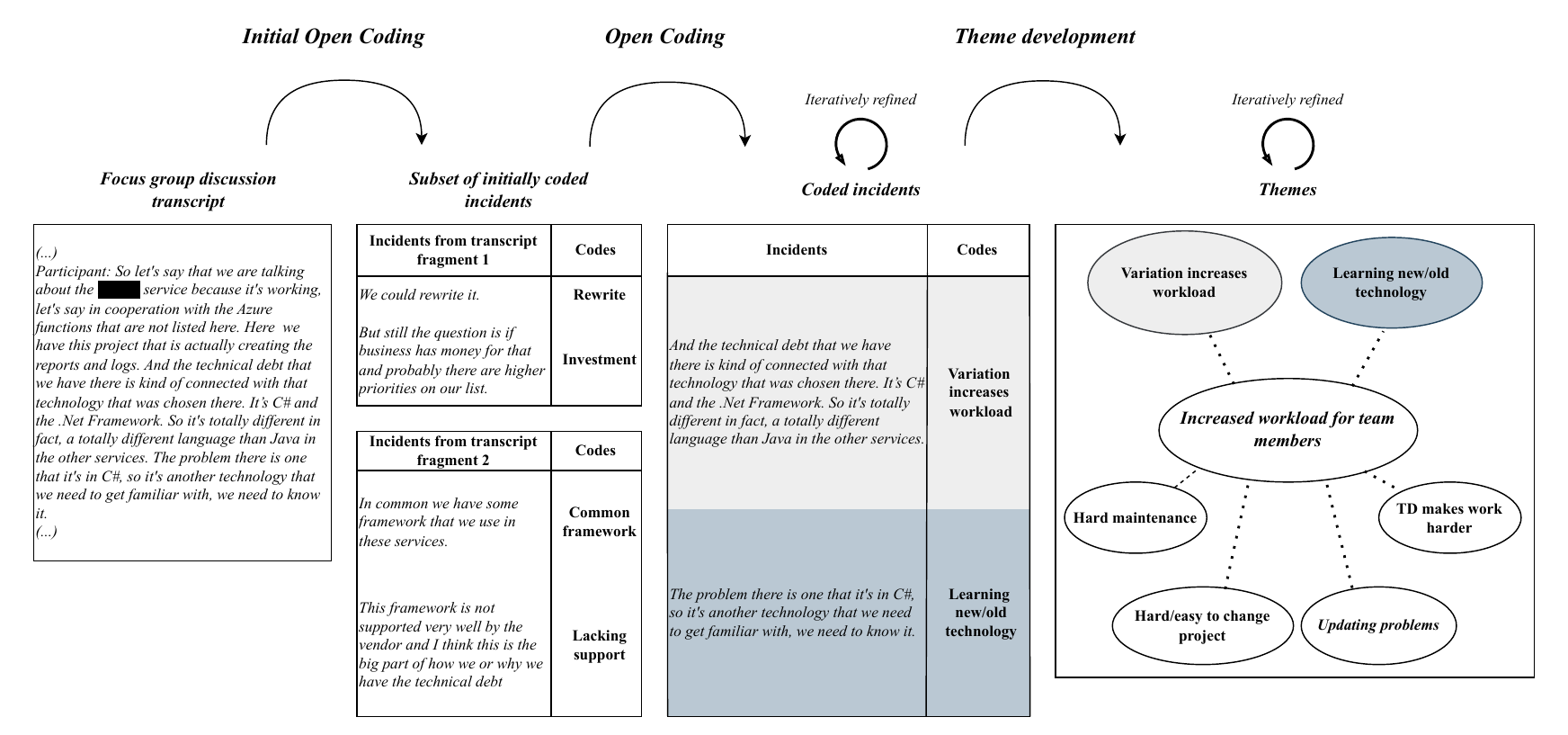}
\caption{Coding process example} \label{fig:coding_example}
\end{figure*}

\subsection{Phase 3: Focus group discussion - data gathering.}
\label{method:focus_group}
In order to gain a complete picture of the TD present in the RS, we complemented the quantitative results with a qualitative analysis. We deemed this process necessary as TD is a liability to internal system quality~\cite{avgeriou2016managing}, which may be hard to measure through a static analysis tool such as SonarQube.

Phase 3 of our research process consisted of a focus group~\cite{kontio2008focus} where the participants were the current development team of the RS (see Table \ref{tab:participants} for development team demographic data). The participants of the focus group were selected by the lead architect of the RS through purposeful sampling, by considering primarily their knowledge of the RS, roles, and availability. They were all part of the team that is responsible for ``foundational microservices'', i.e. the ones that provide core functionalities that are currently the main source of the RS's business value.
This was not the only team that takes part in developing the system - there is also a second team responsible for ``new/experimental microservices'', which we did not interview. This was not due to the researchers' unwillingness to do so, but simply because only the ``foundational'' team was made available to us by our industry partner.

One author moderated the focus group discussion, while two other authors supervised the process and intervened when strictly necessary. The focus group lasted approximately one hour and was transcribed in full for subsequent analysis.

The discussion started with gathering the consent of the participants and ensuring that all the participants possessed sufficient knowledge about the TD concept. Subsequently, participants were asked to point out the microservices that, in their opinion, were most impacted by TD (i.e. the TD hotspots). \tocheck{At this point we did not present our list of TD hotspots, since we wanted to discover which microservices the participants would choose without any suggestions from us.} To aid this process, a de-anonymized version of the diagram depicted in Figure~\ref{fig:plant_uml}, as well as supplementary information on the microservices reported in tabular form, was presented to the participants during the discussion.

For all microservices the participants  picked as TD hotspots, participants were asked a predefined set of 5 questions, namely: 
\begin{enumerate}
    \item ``\textit{On a scale of 0 (none) - 5 (severe), how would you rate the severity \tocheck{of TD in this microservice?}}''
    \item ``\textit{Could you describe the nature \tocheck{of TD in this microservice?}}''
    \item ``\textit{What led to the introduction \tocheck{of TD to this microservice?}}''
    \item ``\textit{Are there any consequences of \tocheck{TD in this microservice?}}''
    \item ``\textit{How are you / will you manage \tocheck{the TD in this microservice?}}''
\end{enumerate}

Such questions were asked in a semi-structured fashion and were used as introductory entry points to gain an as complete as possible understanding of the TD item discussed in follow-up questions. \tocheck{The five questions presented above were designed not to target any specific debt type, e.g., code or architecture TD, leaving this option open to the participants. This premeditated research design choice allowed participants to discuss any TD type, and hence not necessarily focus only the TD types which might be more swiftly detected \textit{via} our source code analysis, e.g., code and architecture debt. As further presented in Section~\ref{sec:qual_results} and Section~\ref{sec:discussion}, this process led to the identification of TD items of different nature in addition to code and architecture TD, such as documentation debt and social debt~\cite{tamburri2013social}.} 
By following the focus group method, all the participants were encouraged to chime in with their perspective at any point during the session, even when TD hotspots picked by another participant were being discussed.

\tocheck{After the participants had thoroughly described the microservices they deemed as TD hotspots, they were presented with the list of potential hotspots identified in Phase 2 of this research} (see Section~\ref{method:potential_TD_identification}). Microservices that were already picked for discussion by the participants were not considered again for discussion during this step.

To ensure that participants were not unconsciously biased in talking about a limited number of TD types, e.g. code or architectural TD, the focus group concluded with an additional step. 
In this final step, for all TD types that were not already mentioned, the participants were asked whether TD of that particular type was present in the RS. The list of TD types was taken from a fairly recent tertiary study collecting TD types from 19 secondary studies on TD~\cite{junior2022consolidating}.  

The transcript of the focus group discussion was first automatically transcribed and then manually reviewed by one of the authors to ensure its fidelity with respect to the original recording. To ensure the replicability and scrutiny of the focus group process, the group discussion plan and slides are available as additional material~\cite{additional_material_jss}.

\subsection{Phase 4: Reflexive thematic analysis.}
 \label{method:reflexive_thematic_analysis}

 Reflexive thematic analysis~\cite{braun2022thematic} is a qualitative research method allowing researchers to find underlying themes in the data. This method is extremely flexible and allows researchers to explore qualitative data in-depth with an openness to new unexpected findings. We employed this analysis method since we wanted to extract the essence of what the development team said during the focus group discussion with regard to their experiences with TD in the RS. 
 
 During this research phase, we followed the guidelines provided by Braun and Clarke~\cite{braun2022thematic}. Specifically, the analysis consisted of the following steps (see Figure \ref{fig:coding_example} for an example):
 \begin{itemize}
     \item Step 1: Pilot of the initial coding \textit{via} open coding~\cite{Saldana2013}, where each of the three authors coded three randomly assigned pages of the transcript. The entirety of the codes were then jointly discussed to establish a common coding style for the open codes.
     \item Step 2: Open coding of the whole transcript by the first author, followed by a coding review from the co-authors. This step was repeated three times before the codes were finalized. During this process, codes were jointly discussed, and initial theme candidates were identified.
     \item Step 3: Codes were grouped together into identified themes, i.e., the central concepts that the codes are related to. This process was done similarly to axial coding, where open codes are grouped around an ``axis'' category~\cite{Saldana2013}.
\end{itemize}

As suggested by Braun and Clarke~\cite{braun2022thematic}, through our reflexive thematic analysis we iteratively analyzed and critically reviewed our ongoing analysis during its iterations. As a result, the themes that ultimately resulted from this process were a product of coding that was cooperatively done by all three authors over a total of five iterations.

Additionally, since we asked the participants to rate the negative impact of TD on each discussed microservice, we noted these ratings through magnitude coding~\cite{Saldana2013} and compared them to the SonarQube Technical Debt measurements. In two cases, the participants explicitly stated an amount of time necessary to repay the technical debt in their opinion, these were also compared to SonarQube measurements.

As a result of this phase, 10 themes were identified, and a comparison between SonarQube results and the participants' TD severity ratings was created.

\subsection{Phase 5: Interview with lead architect}
The last phase's goal was to validate the findings from the qualitative analysis.
In order to do so, we discussed the themes and the comparison of microservices TD severity with the lead architect of the RS in a semi-structured interview. 
Each theme was presented to the architect and feedback was asked regarding the correctness of the findings, any further detail the architect could provide us regarding the themes, and if they were aware of any TD in the RS that was currently missing. 

The final themes were adapted according to the final interview. The entirety of the findings collected with this research, both regarding the quantitative analysis (Phase 1-2) and the qualitative one (Phase 3-5), are presented in the following section.

\begin{figure*}
\center
\includegraphics[width=0.9\textwidth]{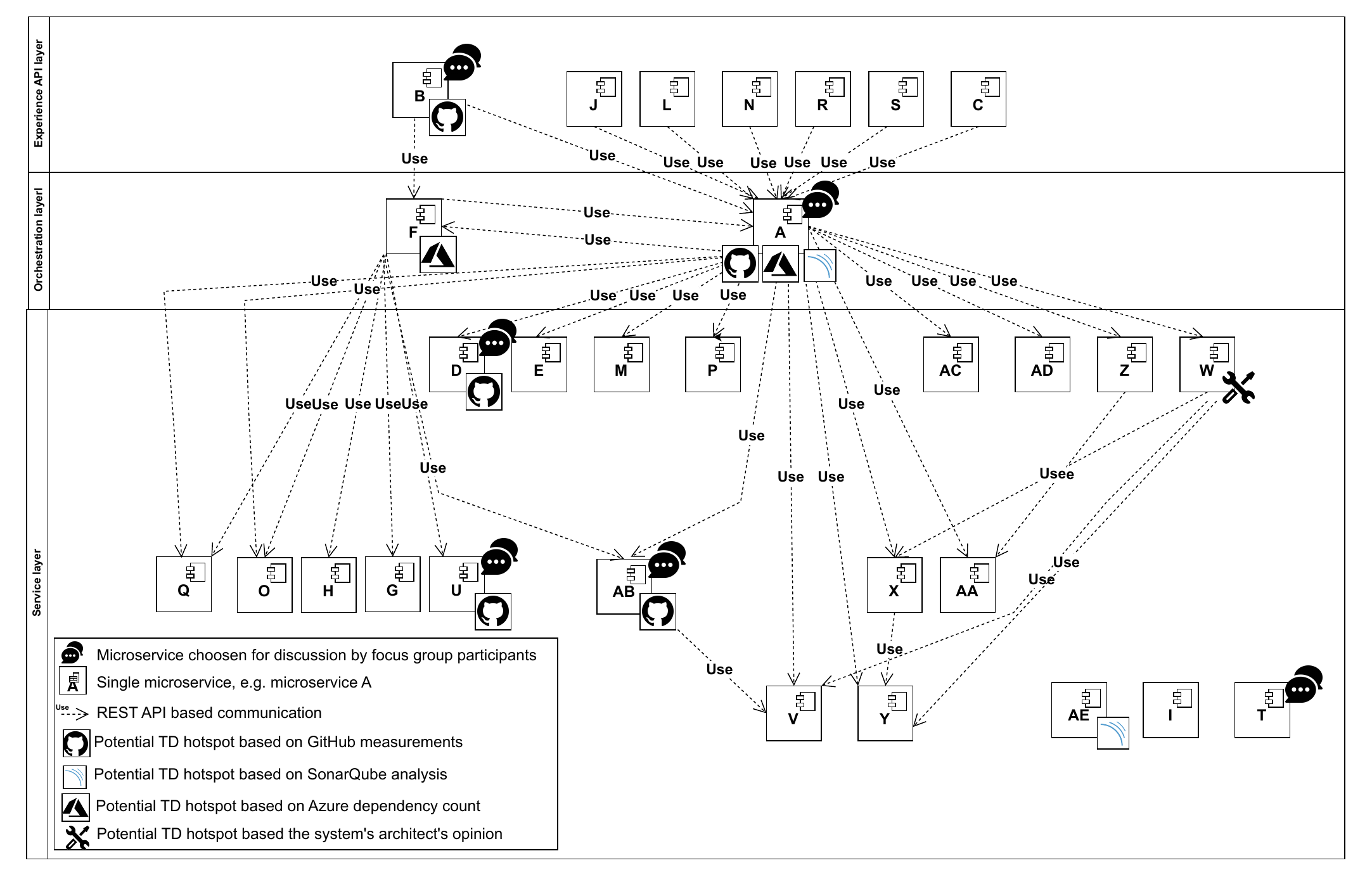}
\caption{Component diagram of dependencies between RS microservices} \label{fig:plant_uml}
\end{figure*}

\section{Results}
\label{sec:results}

\subsection{Quantitative Results}

The reverse-engineered component diagram created through the Azure features (see Section~\ref{sec:research_process}, Phase 1a) is documented in Figure \ref{fig:plant_uml}. The names of the microservices have been anonymized due to a confidentiality requirement from our industrial partner. For simplicity, the microservices are identified in this research by a sequence of letters assigned randomly by following a lexicographical order, ranging from microservice A to microservice AE.

\tocheck{For each microservice, we gathered data from three sources: (1) directly from GitHub repositories, (2) Azure Application insights and (3) SonarQube. 
Basic measures such as lines of code, age and contributor count were obtained directly from GitHub repositories. The number of dependencies was obtained through Azure Application insights, Finally,  SonarQube made it possible to calculate the following: TD (time necessary to repay the TD), TD Ratio (time for TD repayment/time used for developing the code), and TD Density (percentage of code affected by TD). A summary documenting the quantitative measurements, in terms of average, standard deviation, minimum, and maximum values, is reported in Table~\ref{tab:sonarqube}.}
The programming languages used in the system were mainly Java (73.3\% microservices), Typescript (23.3\%), and C\# (3.3\%). Specific measurement values for each microservice are available online for scrutiny as part of our replication package.


\begin{table}
\scriptsize

\begin{tabular}
{llllllll}
\hline
	& LoC&	AGE&	CNT	&	TD	&	TD-R &	TD-D &	DEP	\\
\hline
AVG	&	1683.5	&	2.8	&	15.4	&	906.1min	&	1.9\%	&	0.5	&	2.5	\\
STDEV&	1612.8	&	1.1	&	7.2	&	1237min	&	1.8\%	&	0.5	&	4.4	\\
MIN&	45	&	0.5	&	4	&	10min	&	0.1\%	&	0.02	&	0	\\
MAX	&	6800	&	5	&	37	&	5760min	&	8.5\%	&	2.6	&	24	\\
\hline
\end{tabular}
\caption{\tocheck{Code} analysis summary \\
LoC - Lines of Code,
AGE - Age in years,
CNT - Number of contributors,
TD - Technical debt,
TD-R - TD Ratio,
TD-D - TD Density,
DEP- Number of dependencies}
\label{tab:sonarqube}
\end{table}

From the quantitative data collected, the following microservices were identified as potential TD hotspots, i.e., the microservices that were subsequently discussed in the focus group (see Section~\ref{sec:research_process}):
\begin{itemize}
    \item \textit{Microservice A}: This microservice was the first choice for a potential TD hotspot. It was the central microservice with most dependencies (24), had the highest lines of code count (6800), the most contributors (37), was \tocheck{one of the two equally} oldest microservice (5 years), had the highest level of TD (15 days), and was described by the lead RS architect as ``the fundamental service with most functionalities and logic''. This microservice's role is that of an Orchestrator, which naturally leads to it having the most dependencies.
    \item \textit{Microservice AE}: This microservice presented the highest TD density (2.56\%) and interestingly had no dependencies with other microservices.
    \item \textit{Microservice F}: This microservice was characterized by the second highest number of dependencies (9);
    \item \textit{Microservice B}: While presenting mostly low measurements in terms of lines of code (1000), TD (20 minutes), TD density (0.2\%), and dependencies (2), this microservice was the \tocheck{one of the two equally oldest ones} (5 years). In addition, the RS lead architect mentioned that there had been a recent update issue related to this microservice.
    \item \textit{Microservice D}: This microservice had the second highest lines of code number (5600) and was pointed out by the RS lead architect as the one ``containing the most complex algorithms of RS''.
    \item \textit{Microservice AB}: This microservice was identified as a possible TD hotspot since it was the only one implemented in C\#, which was a programming language not used in any other microservice in the RS.
    \item \textit{Microservice U}: This microservice had the second highest count of contributors~(29).
    \item \textit{Microservice W}: Despite not presenting any outlier  measurement, the RS lead architect mentioned this microservice as a possible TD hotspot, referring to it as ``a proof of concept that stayed.''
\end{itemize}

\subsection{Qualitative Results}
\label{sec:qual_results}

\begin{table*}
\scriptsize

\begin{tabular}
{p{3cm}p{10cm}p{4cm}}
\hline
Theme	& Description & Related Microservices\\
\hline
Obscure inheritance	&	The process of inheriting an existing service from another development team. This includes various problems that come from it: knowledge vaporization due to documentation TD and the need to learn new exotic/obscure technologies and algorithms. & A, AB, B, F, W, Whole System \\
The natural software aging process & The typical process of software aging: Over time, services become bigger and the code more complex, and since there is little refactoring - big services full of code-level TD emerge. & A, AB, B, D, U, W, Whole System \\
Increased workload for team members & Mentions of the additional time/effort that the developers have to do put into their everyday work. 
 & AB, B, D, F, P, U, Whole System \\
 Domino effect possibility & The situation where developers avoid "touching" anything in existing code because they are afraid of unexpected consequences that may affect other components/services.
& A, B, D, U \\
Step by step rewrite strategy & The repayment strategy that the team used to repay ATD - which is rewriting parts of the code in small increments to slowly get rid of TD - usually to get rid of the ``common framework''. & A, AB, D, P \\
Collaboration roadblock & Participants' explanations of why their collaboration with the other development team is lacking. & AE, F, Whole system \\
Responsibility division &Information about the responsibility for specific services/actions/tasks that fall on a specific team. & AE, F, W, Whole system \\
Start-up mode & Incurring TD on purpose in order to fastly develop an MVP that would satisfy business stakeholders & Whole system \\
Developers vs. business  & The conflict between the development team and business stakeholders regarding TD management. Usually, when developers try to obtain time/budget for TD repayment. & AB, P, Whole system\\
"It works" & Mentions that despite various issues with the project, the system/service has the merit of ``Working''/``Getting the job done''. & B, D, W, Whole system \\

\hline
\end{tabular}
\caption{Thematic analysis results overview}
\label{tab:themes_results}
\end{table*}

In this section, we present the themes resulting from the reflexive thematic analysis, which were refined through an interview with the lead architect of the RS (see Section~\ref{sec:research_process}, Phases 3-5). From the emerging themes, characterized by being heterogeneous in nature, we were able to capture a more holistic understanding of the TD present in the RS. An overview of these themes is presented in Table~\ref{tab:themes_results}. 

\textbf{Obscure inheritance.}
The theme ``obscure inheritance'' represents a system that was originally created by people other than the team currently developing and maintaining it. The current team slowly unravels the specifics of this ``inherited'' system, since the knowledge abut the system was not documented appropriately. 

The RS was originally created by a different development team than the current one. This led to numerous problems and challenges, one of them being the vaporization of architectural knowledge. As one participant stated, when asked about the only microservice implemented in C\#: ``\textit{I don't know. I can only imagine that maybe at that time when they were writing it, it was like 3 or 4 four years ago.}''
When asked if anyone working on the RS currently has worked on it since its inception, one participant stated: ``\textit{Definitely there is no one, even the original primary creator of that [the RS].}'' 
As a consequence of lacking documentation, the current team had to recreate the documentation themselves. By putting it in the words of one of the current developers: ``\textit{We perform a lot of archaeology on the services, a lot of investigation to understand what's going on. Based on this, we prepared some documentation\dots and since then we have a fully responsibility for this.}'' 

It is possible that major problems related to the lack of documentation about the architecture could have been at least partially prevented by the use of architectural decision  records~\cite{kopp2018markdown}. However, they were not used during the development of the RS.


\textbf{The natural software aging process.}
The participants often mentioned the classic, widely known, process of software quality declining over time, mostly due to no major effort being invested in keeping a software product up to date as time passes by~\cite{lehman1979understanding} \cite{parnas1994software}.

This theme reflects the typical growth of any piece of software and, in the case of the RS this theme is primarily embodied in the history of ``\textit{microservice A}.''. Firstly, this microservice is the oldest one in the system, as admitted by one participant: ``\textit{It has been created probably as one of the first things in this company}''. Naturally, since it is an old key component, it became relatively large, e.g.``\textit{there are lots of lines of code and many classes}.'' With increased size, the complexity of the logic in it also increased, as another participant described it:``The service is quite complicated.'' 

During this period of growth, there was no refactoring of code that became obsolete and sometimes even unnecessary. For example, participants stated that the service contains a big list of "if" statements related to various customers: `` \textit{IFs for customers that are not with us for months now and no one has fixed that.}''. This ultimately resulted in code that was hardly understandable, as a participant described:``\textit{[\dots] a very big service with spaghetti code with the framework that nobody knows[\dots]}.''

\textbf{Increased workload for team members.}
As commonly experienced and documented \cite{ramavc2022prevalence}, one of the most widespread issues the TD caused in the RS was a notable workload increase.

From the insights provided by the practitioners, the workload increase was mostly due to two seemingly contradictory causes, namely (i) the use of the same technology (a proprietary framework) in most services, and (ii) the use of a specific technology (C\#) in only one microservice. Using a problematic proprietary custom framework in most microservices led developers to face problems when updating the software in many microservices. As described by one participant: ``\textit{I encountered problems with this framework not being forward compatible or backward compatible and stuff like that.}''. On the other hand, using C\# in the case of only one microservice made it necessary to identify developers competent enough to use it in an organization where all other systems were based on either Java or TypeScript. As one participant stated: ``\textit{[\dots] we basically are Java development organization and we hire only Java developers[\dots].So having just one service that is in C\# causes that we need to build or gain somehow the competences [\dots]}.''


The RS lead architect additionally informed us that this increased workload was not only a consequence, but also a cause of TD. Since developers had less time to carefully develop new features due to this increased workload, they incurred more TD, creating a vicious TD cycle\cite{martini2015danger}.

\textbf{Domino effect possibility.}
This theme relates to various dependencies in the system that make developers fearful of touching the existing software, in fear it may cause a ``domino effect'' leading to new problems in unexpected parts of the system.

One piece of key information the participants shared was that there are hidden dependencies that cannot be seen on the component diagram. One of them was that ``microservice AE'' had shared data with other microservices (\textit{``They share the data but they don't share calls or invocations.''}), which means that many other unknown dependencies of a similar nature may exist in the system. Additionally, while separate in functionalities, most microservices used the same proprietary framework. As stated by a participant: \textit{``[\dots] we have a framework that we use in most services.''}. 
Since this proprietary framework was faulty, these seemingly separate microservices shared the same problems. As stated by one participant: \textit{``[\dots] this framework is in the majority of of our services, so it's problem for for the majority of them, for the whole system.''} Ultimately, the team often resorted to the ``no touching'' strategy in the case of possibly common elements, e.g. ``Sometimes it's better not to to touch some stuff''.

The RS lead architect additionally complemented this finding by adding that avoiding this fear of the domino effect was actually the original reason why the system was developed as a microservice architecture rather than a monolithic architecture. While this problem may indeed have been avoided at first, as the RS grew in size, it ultimately did start affecting this large microservice-based system.

\textbf{Step by step rewrite strategy.}
Due to their pressing problems with Architectural TD, the team devised a strategy allowing them to repay TD without having to turn off the production environment of the RS. The strategy can be summarized in four key steps:
\begin{itemize}
    \item Choose the microservice to refactor - the last one in the dataflow that you are trying to refactor. As a participant stated: ``\textit{So this service is kind of at the end of the whole path of the order}.''
    \item Create a clone of the microservice - the same input data should flow through both and the outputs should be compared to check for any differences. As a participant described: ``\textit{the flow would go through both the old and new path and we will compare the results of both version and if there are any differences, we would know about them.}'' The new microservice should not be sending any data to the existing services yet.
    \item In small steps, rewrite pieces of each functionalities and test for any differences in both microservice's behavior. As a participant said: ``[\dots] \textit{we'll be able to monitor how the new version behaves and once we're ready, once we fix some stuff because of course there might be some errors in new version.}''
    \item Switch the new microservice in the place of the old one, when both version's behavior becomes identical. As stated by one participant: ``\textit{Once we're ready, we can switch to the new version and get rid of the old one.}''
\end{itemize}
The participants shared that this approach has allowed the team to replace the proprietary custom framework in the case of \tocheck{microservice T}. 


\textbf{Collaboration roadblock.} 
This theme is centered around various problems that the team participating in the focus group had while collaborating with a second team involved with the system. Overall, three issues making the  collaboration harder and negatively influencing the TD in the RS were pointed out. Firstly, lacking support - the other team took a long time to respond to issues reported regarding their part of the system, e.g. ``\textit{it took like a month, a month for a very trivial thing to be to be fixed, right?}''. Secondly, there were different approaches to establishing best practices, which resulted in inconsistencies in the system. As one participant stated: ``\textit{Let's say they are not following the object oriented principles, they are not following the SOLID rules and so on.}''. Thirdly, shallow communication - the teams did not work closely together, as one participant admitted: ``\textit{[\dots] our contacts are only in case of emergencies or some special cases.}''

\textbf{Responsibility division.} 
This theme represents how the participants' team responsibility differed from the second team's responsibilities. In the case of three microservices that we asked about(F, W, AE) we were not able to collect a satisfactory amount of information, as the microservice was viewed as the responsibility of the other team. This responsibility division was often mentioned by the participants, e.g. ``I think this one might be owned by the other team''. The reason for this divide was that in the focus group participants were responsible for the ``foundational'' services, while the other team focused on new innovative microservices. As one participant described: ``\textit{[\dots] we have two groups of services, one is foundational and one is innovations. Our responsibility is to maintain and develop foundational services, and their is to do good innovation stuff, right.}''

The RS lead architect pointed out that this division of responsibility had one potentially dangerous consequence, which was the emergence of ``orphaned microservices'' that no team felt responsible for, since each team believed that they were the other team's responsibility.


\textbf{Start-up mode.}
The term ``start-up mode'' was used by one of the participants to explain the original sources of lingering TD. It described the situation when a young company had to swiftly deliver an MVP in order to gain enough capital to continue operations. In the participant's words: ``\textit{So the creators were in a rush. They need to deliver value quite quickly.}'' At the time of the RS's early inception, major decisions that resulted in architectural TD were made with the knowledge of possible future consequences. However, the consequences were more severe than expected, as described by a participant ``\textit{We know that up front, let's say, but I think nobody was expecting how hard it would be.}''

\textbf{Developers vs. business.} 
This theme focuses on the conflicting interests of software developers, who wanted ideal code with no TD, with business stakeholders, who prioritized the swift delivery of functionalities that are a source of business value. The participants stated that convincing business stakeholders to allow developers to repay TD in one case was a challenge: ``\textit{it was a really long and bumpy road to convince business to allow us to do it}.'' They also noted that, while TD may not always be visible to business stakeholders, they may face its consequences in the future: ``\textit{[\dots]TD that is not recognized by the business but it will hit them soon.}''

\textbf{``It works''.}
The ``it works'' theme conveys a key idea that, although mentioned only four times, can be considered as extremely important. Despite all of the TD in the RS and all the issues that the team faced, of most importance was the fact that the RS ``worked'' properly and as such, it produced business value. One participant, after stating that ``\textit{There was no documentation at all for this}.'' positively ended with ``[\dots] \textit{the functionality is fully working on production.} [\dots] \textit{we have not had any major issue with this on production.}''

\subsection{\tocheck{Quantitive and Qualitative Results Comparison}}

\tocheck{In order to get more insights into the gathered results, we compare the data collected \textit{via} our quantitative analysis, i.e., the potential TD hotsposts, with the ones gathered through our qualitative analysis. This additional process was conducted in order to understand if, and in affirmative case to what extent, the two research methods adopted were aligned in terms of TD hotspot identification.}

\tocheck{During the focus group discussion participants firstly discussed the microservices of their own choosing in the context of TD. After that, we explicitly asked them about potential TD hotspots that they have not chosen themselves. We did this to avoid biasing participants into discussing only the potential TD hotspots that we obtained through code analysis.}

\tocheck{Additionally, we asked participants to rate the TD in all the discussed microservices on a scale of 0 to 5 (where 5 was the highest value). These values are compared to SonarQube TD measurements in Table \ref{tab:sonarqube_vs_participants}. In the case of three microservices, the participants refused to answer by stating that the microservice is the responsibility of another team. Additionally, a participant provided their own estimate of TD for microservice B by stating that ``\textit{it would take months}'' to repay all the TD that it contains. We noted this estimate since it was significantly higher than any of the SonarQube measurements.}

\tocheck{Figure~\ref{fig:plant_uml} showcases a comparison between microservices chosen in discussion by participants and the TD hotspots that we obtained through our quantitative analysis.
There are exactly four mismatches between these, microservices F, W, AE and T. The mismatch in the cases of microservices F, W and AE were caused by the participants lack of knowledge about these microservices, since a different team was responsible for them. Finally, microservice T was mentioned in the context of being a positive example of an almost ``TD free'' microservice, so we do not consider it as a missed ``TD hotspot''.}

\textbf{\tocheck{GitHub measurements.}} 
\tocheck{The following simple values from the repositories were used to identify TD hotspots: lines of code (microservices A and D), age (microservices A and B), contributors (microservice A and U) and unusual programming languages (microservice AB). All of these microservices were also pointed out by participants during the focus groups as severely impacted by TD, with severity ratings between 2 to 5. In total, all 5 out of 5 TD hotspots identified by Github measurements were confirmed as such by the qualitative analysis.}

\textbf{\tocheck{Azure Insights dependency measurement.}} 
\tocheck{Two microservices were chosen as TD hotspots due to their high dependency count: microservices A and F. Microservice A was confirmed by participants to be severely impacted by TD with a severity rating of 5. Participants lacked knowledge of microservice F due to it being developed by another team. In total, 1 out of 2 TD hotspots identified by Azure Insights dependency count were confirmed by the qualitative analysis, while 1 could neither be confirmed nor disproven.}

\textbf{\tocheck{SonarQube measurements.}}
\tocheck{We choose the following TD hotspots due to SonarQube results: A (TD repayment time), AE (TD density). The Microservice with highest TD repayment time was confirmed as a crucial TD hotspot by participants, with the maximum severity rating of 5. The microservice with the  highest TD density was instead not developed by our participants and they could not confirm nor disconfirm its TD severity. In total, 1 out of 2 TD hotspots identified by SonarQube analysis were confirmed as such by the qualitative analysis, while 1 could neither be confirmed nor disproven.}

\tocheck{However, from the collected results we noticed that the SonarQube estimates often contradicted the level of TD perceived by the participants. This is exemplified by the comparison of microservices T and B. 
Microservice T was mentioned by participants specifically to showcase a microservice that they believed to be the most ``TD free'', this is a young microservice (0.5 years) where the team intentionally avoided incurring the architectural TD that was plaguing most microservices in the system, namely the common proprietary framework. As such, it received a TD severity ranking of 1 from participants, despite a high SonarQube TD estimate (2 hours and 9 minutes).
In the case of microservice B, participants gave an estimation of ``months'' for TD repayment and gave a TD severity ranking of 3. However, the SonarQube's TD measurement was very small - only 20 minutes.}

\tocheck{\textbf{Comparison conclusion.}}
\tocheck{Overall, participants have confirmed 5 out of 8 TD hotspots identified by our simple code analysis. The 3 TD hotspots that were not conformed, were related to microservices that the participants had no knowledge about. As such, all potential TD hotspots that could be confirmed, were found to be relevant to identifying major TD items. Therefore, it seems that the simple code analysis is a good entry point for holistic TD discovery. }



\begin{table}
\scriptsize

\begin{tabular}{lllll}
\hline

Severity 	&	Microservice& TD& TD Ratio & TD Density\\
\hline
1	&	T	&	2h9min	&	0.3\%	&	0.1	\\
3	&	B	&	20min	&	0.2\%	&	0.02	\\
4	&	D	&	4d1h	&	1.2\%	&	0.35	\\
4	&	AB	&	10min	&	0.7\%	&	0.22	\\
5	&	A	&	12d	&	3\%	&	0.85	\\
2 or 3	&	U	&	1d2h	&	1.7\%	&	0.46	\\
Unknown	&	F	&	3d3h	&	1.5\%	&	0.44	\\
Unknown	&	W	&	3d6h	&	2.4\%	&	0.73	\\
Unknown	&	AE	&	3d6h	&	8.5\%	&	2.56	\\
\hline
\end{tabular}
\caption{SonarQube measures comparison to participants' TD \\severity~perception
}
\label{tab:sonarqube_vs_participants}
\end{table}


\section{Discussion}
\label{sec:discussion}
We analyzed the TD in the RS both through a qualitative lens by utilizing data provided by Azure, \tocheck{GitHub} and SonarQube, and refined the inquiry results \textit{via} a qualitative approach based on a focus group and interview with the lead architect. From the results collected in this case study, we can derive several implications and takeaways, which are discussed further in this section.

Firstly, we observed that SonarQube's estimates of the time required to repay TD were very different from the team members' TD estimates. This finding corroborates and builds upon existing findings on SonarQube remediation times~\cite{baldassarre2020diffuseness}.In the considered case study, the SonarQube estimates turned out not to be representative due to the fact that the tool could not detect TD \tocheck{in its entirety}. In fact, the main architectural TD item that was impacting the system, namely a proprietary framework used in most microservices, was not detected and, to the best of our knowledge, is not currently detectable \textit{via} static or dynamic source code analysis. However, it is worth noting that most microservices that we hypothesized to be potential TD hotpots \textit{via} \tocheck{a combination of simple static metrics provided by GitHub, Azure, and} SonarQube did match the participant's view on the most TD-impacted microservices. In fact, all of the mismatches were indicated by the participants as \tocheck{microservices that were the} responsibility of a different team, i.e., \tocheck{perticipants did not possess} the knowledge on TD severity in these microservices. This indicates that the \tocheck{combination of simple static analysis} measurements can be a good starting point for a comprehensive mixed-method TD analysis, through it would necessarily require an additional qualitative input from the development team to be regarded as complete.

\begin{highlight}
    \textbf{Take-away 1: \tocheck{Simple source code static analyses} can be an entry point for holistic \tocheck{mixed-method} TD discovery.} When used in isolation, simple static analyses may result in the omission of key TD items. However, \tocheck{pointers from simple source code analyses (e.g, GitHub metrics, microservices interdependencies, and SonarQube TD values)} can be successfully used to pinpoint which microservices may be more impacted by TD and therefore require further qualitative, analysis. 
\end{highlight}

Three prominent themes arising from the qualitative analysis of TD in the case study, namely ``collaboration roadblock'', ``responsibility division'' and ``developers vs. business'', were characterized by the common axis ``communication''. In the case of these first two themes, communication was lacking between the development teams. Since teams were responsible for different microservices, possessed different working-styles, and rarely contacted each other, this led to vicious collaboration cycles such as finger-pointing frictions, followed by orphaned microservices, and TD increase. This means that, at some point, neither team felt responsible for certain microservices (which they believed was the other team's responsibility) and as such the quality of these microservices dropped and the knowledge about the orphaned microservices vaporized. In the case of the ``developers vs. business'' theme instead, the communication between the development teams and business stakeholders was the main concern. TD management, and in particular TD repayment, requires resources that must often be negotiated with business stakeholders, who are usually unwilling to invest in TD repayment since it does not directly provide business value, and is hard to measure.

\begin{highlight}
    \textbf{Take-away 2: Fostering communication to mitigate inadvertent TD.} A~primary cause of inadvertent TD in a large-scale microservice-based system can be lack of communication. Communication issues may arise either across development teams (resulting in ``orphaned microservices'' and TD increase) or between developers and management (resulting in a lack of time for refactoring). Clear communication about TD-related actions and responsibilities across teams and management can be a crucial factor to prevent inadvertent TD in microservice-based systems.
\end{highlight}

From our case study, it emerged that the original idea behind using a microservice architecture (rather than a monolith) in the system was to avoid tight coupling and thus make it possible for developers to be less afraid of refactoring. However, this approach was ultimately ineffective. While the system grew in size, the complexity and the amount of dependencies in the system grew in an uncontrolled fashion. 
\tocheck{This result emerged in our study \textit{via} our mixed-method approach, where quantitative data, collected using simple static analysis tools such as GitHub, Azure, and SonarQube, were complemented with qualitative insights gathered from focus groups and interviews. This mixed-method approach led to a deeper understanding of the technical debt~(TD) present in the system, which might not have been easily identified by solely considering in isolation a source code analysis or a purely qualitative approach. For example, pointers from the source code analyses supported the identification, \textit{via} the subsequent qualitative process, of a social debt item~\cite{tamburri2013social} manifesting as collaboration frictions between two geographically distant teams.}
As emerging from a \textit{post mortem} reflection of the RS architect following our analysis, despite optimistic expectations, frictions between the organizational and architectural structures led to the emergence of novel architectural issues. In perfect line with Conway's law,  the misalignment between the architecture and  the communication structure, the inter- and intra-teams \textit{modus operandi}, and the geospatial distribution of the organization played a key role in the introduction of inadvertent TD in the case study system.

\tocheck{As the final discussion takeaway on the misalignment with Conway's law observed in the studied case, we speculate that, due to their inherent architectural complexity, microservice-based systems may be more inclined to incur social debt when compared to other architectures, e.g., monolithic ones. Specifically, the modular nature of microservice architectures, combined with intricacies arising from hidden microservice independencies, may foster conditions where social debt items such as miscommunication, friction, and working in isolation among teams assigned to different microservices are more prone to prosper.}

\begin{highlight}
    \textbf{Take-away 3: Misalignment between architectural and organizational structures can influence TD.} Following Conway's law, misalignments between communication structures, inter- and intra-team \textit{modus operandi}, and development culture with the architecture of a system can lead to the natural emergence of TD items. \tocheck{Given their architectural complexity, miscorservice-based systems may be more prone to incur social debt.}
\end{highlight}

\begin{figure}
\center
\includegraphics[width=0.8\columnwidth]{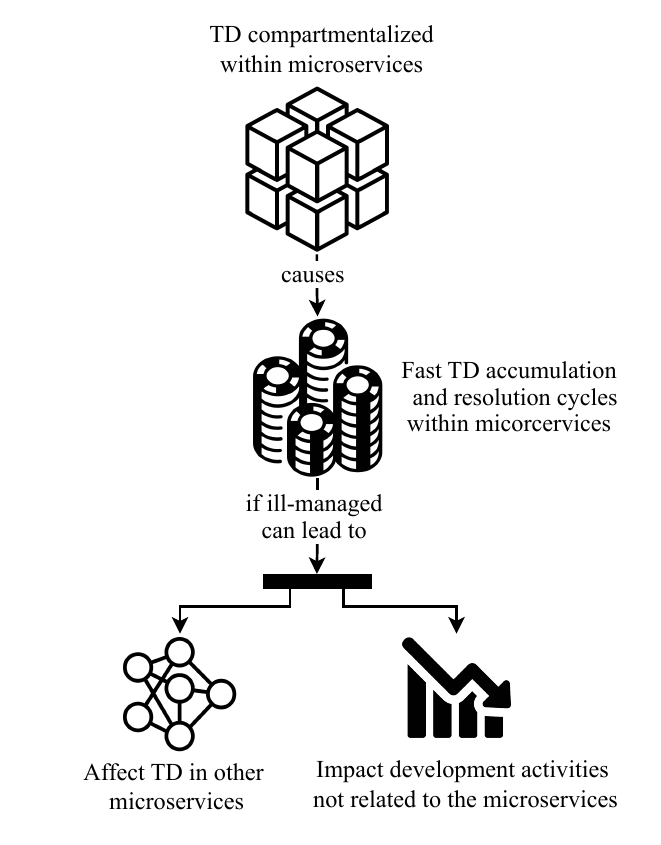}
\vspace{-15pt}
\caption{Overview of the key components of the \textit{microservice architecture technical debt gamble}}
\label{fig:TD_gamble}
\end{figure}

From the quantitative and qualitative analysis, a phenomenon potentially characteristic of MSAs emerged. An intuitive depiction of this phenomenon, which we refer to as ``\textit{MSA TD gamble}'' from now on, is depicted in Figure~\ref{fig:TD_gamble}.

\tocheck{In the considered case study, TD resulted to be perceived as compartmentalized within each  single microservices. Such TD compartmentalization could be attributed to a correct adherence to the MSA pattern, which advocates for loosely coupled microservices, hence isolating TD at the microservice level. Due to perceived compartmentalization of TD within each microservice however, developers felt confident in rapidly accumulating TD within microservices as, due to its independence from other microservices, such TD was perceived as swiftly resolvable in isolation. We refer to such fast TD accumulation and resolution cycles within isolated microservices as the ``\textit{TD gamble}''.}

\tocheck{The risk implied by the observed TD gamble is that, given the rapid accumulation of TD within microservices due to its perceived effortless repayment, the TD in a microservice can also swiftly become unmanageable. Losing the TD gamble, i.e., accumulating an unmanageable amount of debt within a microservice, can lead to inadvertently spread the TD to other microservices due to growing hidden dependencies (e.g., hard-coded inter-microservice calls leading to the ``Domino effect possibility'', see Section~\ref{sec:qual_results}) or more generally affect other development activities due to the effort needed to refactor the microservice originally affected by high amounts of TD.}

\tocheck{As described by one developer during the focus group: ``\textit{We had a colleague who started working in our project and he was given a task to try rewrite the whole service and make it simpler and actually he failed\dots}''}



\begin{highlight}
    \tocheck{\textbf{Take-away 4: The \textit{microservice architecture technical debt gamble}.} Due to the perceived TD compartmentalization within single microservices, TD at microservice level can experience fast accumulation and resolution cycles. Swift accumulations of TD however can easily become unmanageable, leading to inadvertently affect TD in other microservices or impact negatively general development activities.}
\end{highlight}



Despite all the identified issues related to TD, to date, the RS is a commercial success that keeps providing high business value to the partner company.

At the time of data gathering, the development team was making their first steps towards TD management. The development team already recognized that the system had reached a point when a switch from ``start-up mode'', when TD was heavily incurred, to ``long-term quality mode'', was necessary and would require active TD management. To do so, they had developed their ``step by step rewrite strategy'' for TD repayment and took part in this study to identify major TD items and strategies for TD management.

This study allowed the partner company to understand how TD in the RS has grown inadvertently out of control due to the \textit{MSA TD gamble}. As a result, in addition to the ``step by step rewrite strategy'', they decided that the following should be done: TD monitoring, systematic incremental efforts guided by the motto of the boy scout rule~\cite{verdecchia2021building}, improving communication and collaboration across development teams with particular focus on responsibility division to avoid orphaned microservices, improving communication about TD with business stakeholders, and maintaining an up-to-date comprehensive architectural overview by the architect.

\begin{highlight}
    \textbf{Take-away 5: Conscious TD management is as crucial for long-term MSA success.} Notable care is needed to prevent TD from rapidly growing out of control in an MSA. Key TD management strategies include continuous TD monitoring, the ``step by step rewrite strategy'', incremental improvements guided by the "boy scout rule", effective cross-team communication, and maintaining an up-to-date architectural overview.
\end{highlight}

\section{Threats to Validity} 
\label{sec:threats_to_validity}
In this section we discuss the primary threats to validity of this research, by following the categorization of case study threats by Runeson~\cite{runeson2009guidelines} and considering common shortcomings of threats analysis~\cite{verdecchia2023threats}.

\textbf{Construct Validity.}
An inherent construct validity of this case study lies in the shared definition among researchers and practitioners of the TD phenomenon. To mitigate this threat, we relied on the Daghstul 16162 definition of TD~\cite{avgeriou2016managing}, and ensured all the participants shared a common interpretation of TD at the beginning of the focus group. The reductive focus on code TD by 
SonarQube (see Section~\ref{sec:research_process}) was mitigated by complementing the code analysis results with two qualitative processes, namely a focus group and an interview with the lead developer of the case study system. Finally, the last threat that should be considered is that we could not include the feedback of the second development team of the RS in this research, which may have provided additional insights.

\textbf{Internal Validity.}
To mitigate the impact of confounding factors on our results, we (i) utilized the default software analysis tool configuration settings as currently used in the company, (ii) followed focus group guidelines to ensure all participants were able to provide insights~\cite{kontio2008focus}, and (iii) adopted a multi-round reflexive thematic analysis process involving three researchers to analyze the focus group data.

\textbf{External Validity.}
Due to the nature of case study research, the reported results are intended to enable analytical generalization, but should not be considered as statistically representative. Intuitively, results such as \textit{obscure inheritance}, \textit{domino effect possibility}, and \textit{collaboration roadblock} (see Section~\ref{sec:results}) may extend to other microservice-based systems. However, further research would need to be conducted to understand whether and to what extent the collected results are generalizable.

\textbf{Reliability.}
While abiding by the non-disclosure agreement with our partner and ethical guidelines governing this study, for scrutiny and replication purposes, we have made available an anonymized version of all the quantitative results collected for this study, as well as the complete material used for the focus group~\cite{additional_material_jss}. The focus group transcript is not made available to preserve the confidentiality of our industrial partner and protect the anonymity of all the participants partaking in the study.

\section{Conclusion}
\label{sec:conclusion_and_future_work}
In this study, we analyzed TD in a large-scale industrial software-intensive system that comprises a total of over 100 microservices, and is currently in use in over 15k locations.

We utilized a mixed-method industrial case study that relied on quantitative methods (architecture reverse-engineering \textit{via} Azure DevOps tools and TD analysis \textit{via} SonarQube) and a qualitative approach (focus group discussion refined \textit{via} an interview with the lead architect of the system).

Through this study, we managed to obtain a clear picture of the TD residing in the case study product, in particular, from the most pressing TD items to the main TD sources and consequences. At the time of conducting this study, while armed with the knowledge collected through this analysis, the development team is also facing the challenge of shifting from ``start-up mode'' (i.e., delivering value fast while incurring in TD) into a more sustainable ``long-term mode'' of maintenance and development. 

This study made the development team understand how the \textit{MSA TD gamble} had influenced the RS so far and that their strategy of TD repayment through the ``step by step small rewrites,'' while successful so far, was insufficient in terms of long-term TD management.

\subsection{\tocheck{Current works}}
\tocheck{Currently,} there is an ongoing effort to improve TD management in the partner company. The most pressing issue is improving communication, both between the development teams and between developers and business stakeholders. The development teams need to clearly define responsibilities and collaborate closely to minimize the impact of the \textit{MSA TD gamble} and to avoid orphaned microservices. 
Not addressing this impediment could be crucial, e.g., when considering the scenario in which a key microservice stops working properly, and nobody has knowledge about it. To avoid this, discrete and moderated consultations across teams are needed to clarify the responsibilities of each microservice, jointly with discussions on how the responsibilities would be changed, managed, and documented in the future.
Additionally, the development team must clearly communicate with the business stakeholders to make them understand that investing in TD management is necessary and how it would be beneficial in the long-term.

Furthermore, the TD identification performed during this study made the partner company realize that ongoing TD monitoring and maintaining an up-to-date architectural overview are necessary for TD management and communication about TD. As such, the partner plans to implement regular TD monitoring efforts.

\subsection{\tocheck{Future work}}
Future work could include: (i) supporting the team in resolving the TD detected \textit{via} this inquiry and in implementing long-term TD management strategies, and (ii) running an additional specialized analysis of TD in the case study system at an architectural level~\cite{fontana2017arcan} \cite{ospina2021atdx}.

\bibliographystyle{elsarticle-num} 
\bibliography{bibliography.bib}






\end{document}